\documentclass[12pt]{iopart}
\usepackage{iopams}
\begin{document}
\title[Scalar field entanglement entropy for small Schwarzschild black hole]
{Scalar field entanglement entropy for small Schwarzschild black
hole}
\author{M V Teslyk and O M Teslyk}
\date{}
\address{Taras Shevchenko Kyiv
University, Physics Department, 2 Glushkov Prosp., Build. 1, 03680
Kyiv, Ukraine} \ead{machur@ukr.net}
\begin{abstract}
We consider scalar field entanglement entropy generated with black
hole of (sub)planck mass scale thus implying the unitary evolution
of gravity. The dependence on the dimension of the Hilbert space for
degrees of freedom located behind the horizon is taken into account.
The obtained results contain polylogarithmic terms.
\end{abstract}
 \submitto{\CQG}
\section{Introduction}
Bekenstein demonstrated in 1973 \cite{bek73} that black hole (BH)
entropy $S_{\rm BH}$ is proportional to its horizon area $A$. As it
is well-known from the `no-hair' theorem the observer can determine
the mass, angular momentum and the charge of the BH but no other
properties thus giving rise to the entropy origin problem. The
second problem is the information loss problem initiated by Hawking
in 1975 \cite{haw75} that in combination with \cite{bch73}
determined that the BH entropy $S_{\rm BH}=A/4$ in Planck units. BH
appeared to generate a large amount of entropy with no chance to
read out any information almost about the matter falling below its
horizon thus implying the non-unitary evolution. To date solution
for the both problems is a challenging task for any theory of
quantum gravity. At present many approaches to answer them have been
proposed.

In 1996 Strominger and Vafa \cite{sv96} proposed to consider BH
entropy with the help of string theory. Such an approach has
appeared to be a fruitful one; for more information on the topic one
can read reviews such as \cite{hor07} or the recent ones
\cite{pre10, gom11} and references therein.

Another approach to the problems is based on the loop quantum
gravity. It provides counting of microscopic BH microstates and
therefore determinates its entropy. One can read more on the topic
in \cite{khr04, khr05, jac07} and in recent papers such as
\cite{bia09, lt12}. In \cite{sl10} authors applying the similar
approach conclude that BH radiation spectrum should become less
entropic as it evaporates. It leads to the possibility of
information recovery from the BH due to the increasing role of
quantum effects in the evolution of small BH. Such a conclusion
witnesses in favor of information preservation with BH that was
reconsidered first by Hawking in 2005 \cite{haw05} and may interfere
with our own results presented in \cite{bt09}.

Estimation of BH entropy with the help of quantum tunneling approach
is proposed in such papers as \cite{ran11}.

In spite of variety of methods to calculate $S_{\rm BH}$ it is
widely accepted that the entropy should be generated with the BH
event horizon. As the horizon separates the whole spacetime into
accessible and non-observable regions any distant observer should
trace out all the degrees of freedom localized inside the BH.
Therefore the horizon can serve not only as an entropy generator,
but as a depository for the degrees of freedom giving rise to it.
Such an approach was proposed with Srednicki \cite{sre93}; for the
review see \cite{dss08}. Investigation of the BH horizon as the
depository results in the holographic principle \cite{sus95}; among
all the publications on the topic we would like to mention
\cite{rt061, rt062} where the authors use holography to calculate
the entanglement entropy from AdS/CFT.

In \cite{bk11} energy and entropy divergences arising in 't Hooft's
brick wall model \cite{hoo85} are considered in the framework of the
uncertainty principle. The authors raise the question of similarity
between the entanglement and the statistical definitions of BH
entropy. In \cite{kk12} higher order corrections within the brick
wall formalism for arbitrary spin have been found.

As the horizon separates the space into observable and
non-observable one, it is logical to assume that $S_{\rm BH}$
originates from entanglement. Such a viewpoint is interesting since
then there is no information loss at all and gravity obeys unitary
evolution \cite{haw05}. For the other approaches resulting in the
unitary nature of the BH evolution one can read \cite{zcyz09} and
recent review \cite{sol11}. The situation has much in common with
the restricted access to some code represented in the Schmidt basis:
being able to read some part of the code only, one concludes in
non-zeroth entropy. But reducing the inaccessible part of the code
reduces the entropy; the similar behavior should be observed for the
decreasing horizon area. Such a process (i.e. accessing the part of
the code only) can not be viewed as the one that generates entropy.

In this manuscript we apply the approach presented in \cite{bt09} to
BH of (sub)planck mass scale. We count the entanglement entropy of
scalar field separated with the BH's horizon into two parts and
compare it to $S_{\rm BH}$. We realize that the case of the planck
scale BH is speculative a bit. As it is mentioned in \cite{haw75}
such small BHs can not be considered as some classical background
metric for any quantum field. In such a case, one is expected to
apply quantum gravity. But, as such a theory has not been built till
now we try to take a look beyond.

On author's opinion, the (sub)planck mass scale BHs seem to be of
great interest in the light of modern heavy ion collision
experiments. In case of appearance during the LHC experiments small
BHs will witness other dimensions via their spectrum radiation
characteristics, and therefore we hope our investigation might be
useful in the sphere. Also the results we derived may be helpful in
the analysis of quark-gluon bag models.

Our approach has much in common with the others presented in
\cite{ilv02, ilv04, jp07}. However our approach has some
differences. We estimate entropy via volume of BH and angular
momentum while in \cite{ilv02, ilv04} shell volume near the horizon
and momentum are utilized. As for \cite{jp07} it is based on the
thermal atmosphere surrounding the horizon.

This paper is organized as follows. In \sref{concepts} we present
general idea of our model. \Sref{model} discusses the model itself.
Entropy estimation and its analysis are presented in
\sref{estimation}. Discussion and conclusions one can read in
\sref{discussion}.

\section{Basic concepts}\label{concepts}
Throughout the paper Planck units were used.

We consider Schwarzschild black hole of mass $M$ and consequently of
radius $r=2M$ and some scalar field surrounding it. The field is
supposed to be in some pure state $|\Phi\rangle$ in the Kruskal
frame of reference (FR) and to have no influence on the background
metric (quasiclassical approach). Such a condition implies that the
field energy is negligible comparatively to BH's mass.

In the following one should keep in mind that vacuum concept is not
invariant with respect to FR, as it was shown by Unruh \cite{u76}.
Hence, observers in different FRs will detect different states of
the scalar field; this is the crucial point in this paper.

Observer in Kruskal FR will not detect the BH's event horizon. At
the same time, observer from the accelerated, i.e. from the
Schwarzschild, FR will detect the horizon with the particles being
radiated with it. The creation and annihilation boson operators in
both FRs are connected via the Bogoljubov transformations
\cite{haw75,u76}
\begin{eqnarray*}
a^{\dag}=\frac{1}{\sqrt{1-\zeta^2}}b_{\rm
out}^{\dag}-\frac{\zeta}{\sqrt{1-\zeta^2}}b_{\rm in},\qquad
a=\frac{1}{\sqrt{1-\zeta^2}}b_{\rm
out}-\frac{\zeta}{\sqrt{1-\zeta^2}}b_{\rm in}^{\dag},
\end{eqnarray*}
where $a,a^{\dag}$ are the annihilation and creation operators in
the Kruskal FR, $b_{\rm in(out)},b_{\rm in(out)}^{\dag}$ are the
annihilation and creation operators in the accelerated FR inside
(outside) the horizon, and $\zeta$ is defined as
\begin{eqnarray}\label{zeta}
\zeta=\exp\left(-4\pi M\omega\right),
\end{eqnarray}
where $\omega$ is the energy of the field quanta generated at the
BH's event horizon under the Unruh effect.

The Kruskal field $|\Phi\rangle$ will be detected with the observer
from the accelerated FR in the state
\begin{eqnarray}\label{Kruskal phi}
|\Phi\rangle=\sqrt{\frac{1-\zeta^2}{1-\zeta^{2N}}}\sum_{n=0}^{N-1}\zeta^n|n\rangle_{\rm
in}|n\rangle_{\rm out},\qquad a|\Phi\rangle=0,
\end{eqnarray}
where $N=N_{\rm in(out)}$ is the dimension of the in- (out-) Hilbert
subspaces \cite{u76}. Here the in- and out-components (denoted by
the corresponding subscripts) describe the parts of the field under
and above the horizon. We emphasize that now we are working with a
single mode of the scalar field only. Later we will integrate over
all $\omega$ possible to take into account all the modes.

As one can see, \eref{Kruskal phi} is exactly the Schmidt
decomposition \cite{sc07,ek95}, and hence one obtains density
matrices of the in- and out-components
\begin{eqnarray*}
\eqalign{\rho_{\rm in} = \Tr_{\rm
out}|\Phi\rangle\langle\Phi|=\frac{1-\zeta^2}{1-\zeta^{2N}}\sum_{n=0}^{N-1}\zeta^{2n}|n\rangle_{\rm
in}\langle n|,\\
\rho_{\rm out} = \Tr_{\rm
in}|\Phi\rangle\langle\Phi|=\frac{1-\zeta^2}{1-\zeta^{2N}}\sum_{n=0}^{N-1}\zeta^{2n}|n\rangle_{\rm
out}\langle n|.}
\end{eqnarray*}

As we see, different observers handle different density matrices.
Though the Kruskal observer detects the pure state $|\Phi\rangle$,
the accelerated one, because of having access to the out-component
of the field (i.e. to the outgoing radiation) only, detects a
mixture with entropy
\begin{eqnarray}\label{sigma}
\sigma\left(N,\zeta\right)&=-\Tr\,\rho_{\rm out}\ln\rho_{\rm
out}=-\frac{1-\zeta^2}{1-\zeta^{2N}}\sum_{n=0}^{N-1}\zeta^{2n}
\ln{\left(\frac{1-\zeta^2}{1-\zeta^{2N}}\zeta^{2n}\right)}\nonumber\\
 &=-\ln\frac{1-\zeta^2}{1-\zeta^{2N}}
-\left(\frac{\zeta^2}{1-\zeta^2}-N\frac{\zeta^{2N}}{1-\zeta^{2N}}\right)\ln\zeta^2,
\end{eqnarray}
where the following relation
\begin{eqnarray*}
\sum_{n=0}^{N-1}n\zeta^{2n}=\frac{1}{2\ln\zeta}\partial_\alpha\left.\sum_{n=0}^{N-1}\zeta^{2n\alpha}\right|_{\alpha=1}
=\frac{\left(1-\zeta^{2N}\right)\zeta^2-N\left(1-\zeta^2\right)\zeta^{2N}}{(1-\zeta^2)^2}
\end{eqnarray*}
has been used.

As one can see from \eref{sigma}, $\sigma\left(N,\zeta\right)$
depends on 2 parameters: $N$ and $\zeta$. The main problem here is
to estimate the value of $N$. It is easy in the asymptotic of large
BH, as one can use the limit $N\to\infty$ then; such an asymptotic
is popular in the literature. In \cite{bt09} it was applied too.
However, here we consider small BHs with mass $M\leq1$ and therefore
have to take into account finiteness of $N$. Direct estimation of
the magnitude of $N$ may be done via calculation of the field
energy. However it leads to the integral which can be solved
approximately only and therefore is not discussed here. Anyway we
expect $N\gg1$ since otherwise the scalar field will influence the
BH and thus violate the quasiclassical approach. Such an assumption
seems to be reasonable and to have no contradictions with the model.
For BH with (sub)planck mass $M\leq1$ and for the small rest mass of
scalar field quanta $N$ must be large enough to encode all the
degrees of freedom.

\section{Model construction}\label{model}
Expression \eref{sigma} is written for some mode with fixed
parameters of the radiated field component. Model construction
requires correct contribution estimation of all the modes to the
entropy, that is of the system symmetry and of the energy spectrum
determined with $\omega$.

Due to the spherical symmetry we must take into account the
contributions from all the angular momenta $l$ and its projections
$-l\leq\mu\leq l$ possible. The range on $l$ is well-defined and can
be written in the following form:
\begin{eqnarray*}
0\leq\sqrt{l(l+1)}\leq\sqrt{L(L+1)}=rp=2M\sqrt{\omega^2-m^2},
\end{eqnarray*}
where $p$ is the momentum of the field quantum radiated away.

In such a case the entropy from \eref{sigma} should be multiplied
with
\begin{eqnarray}\label{l contr}
\sum_{l=0}^{l=L}\sum_{\mu=-l}^{\mu=l}1=
4M^2\left(\omega^2-m^2\right)+\frac{\sqrt{16M^2\left(\omega^2-m^2\right)+1}+1}{2}.
\end{eqnarray}

But, taking into account angular degrees of freedom is not enough.
In \eref{sigma} $\sigma\left(N,\zeta\right)$ is defined for the
fixed $\omega$ only. As we want to estimate the contribution from
all the modes we should integrate over all the $\omega$ possible.
Therefore we write down the following integral for the radiation
entropy $S\left(N,M,m\right)$:
\begin{eqnarray*}
S\left(N,M,m\right)=\frac{V}{(2\pi)^3}\int_m^M\sum_{l=0}^{l=L}\sum_{\mu=-l}^{\mu=l}\sigma\left(N,\zeta\right){\rm
d}\omega,
\end{eqnarray*}
where $V=4\pi r^3/3=2^5\pi M^3/3$ is the BH volume confined with the
horizon and $m$ is the rest mass of the radiated quanta. The upper
integral bound is equal to $M$ here because the energy of the field
quanta can not exceed BH mass. Substituting \eref{sigma} and \eref{l
contr} into the integral we obtain for the radiation entropy
\begin{eqnarray}\label{S}
\fl\frac{S\left(N,M,m\right)}{S_{\rm BH}}=
\frac{M}{6\pi^3}\int_m^M\sigma\left(N,\zeta\right)
\left[1+8M^2\left(\omega^2-m^2\right)+\sqrt{1+16M^2\left(\omega^2-m^2\right)}\,\right]{\rm
d}\omega,
\end{eqnarray}
where $S_{\rm BH}=4\pi M^2$ is the Bekenstein-Hawking entropy and
$\zeta$ is defined in \eref{zeta}.

\section{Entropy estimation}\label{estimation}
Before we proceed, let us make some estimations of $m$. In case of
being not equal to 0 $m$ should be of elementary particles mass
order, i.e.
\begin{eqnarray}\label{m bounds}
m=0 \qquad\mbox{or}\qquad10^{-23}\leq
m\leq10^{-18}\qquad\Rightarrow\qquad m\gtrapprox0,
\end{eqnarray}
where $10^{-23}$ is of order of the electron mass $m_e$ and
$10^{-18}$ is of order of the Z$^0$ boson mass $m_{Z^0}$. So it
should be taken into account that $m$ is a small number.

The integral \eref{S} can not be calculated directly due to the
strong integrand dependence on the integral boundaries. The
situation is complicated with the exact entropy dependence on $N$.
In \cite{bt09} the dependence on $N$ had been neglected because of
the large BH mass, but here we can not do the same trick.

To estimate the entropy at first we decompose
$\sigma\left(N,\zeta\right)$ from \eref{sigma} into series
\begin{eqnarray}\label{sigma ser}
\sigma\left(N,\zeta\right)=\sum_{n=1}^{N}\zeta^{2n}\left(\frac{1}{n}-2\ln{\zeta}\right)
-N\zeta^{2N}\left(\frac{1}{N}-2\ln{\zeta}\right)+\mathcal{O}\left[\zeta^{2(N+1)}\right].
\end{eqnarray}
Due to \eref{zeta} $\zeta$ exponentially depends on $\omega$, and
thus such a decomposition is good at the higher integral bound
$\omega=M$. But one can argue that at the lower integral bound
$\omega=m$ such a series expansion may fail: due to smallness of
$m$, that follows from \eref{m bounds}, $\zeta$ will not differ from
unity a lot. However the neglected terms in the expansion are of the
order $\zeta^{2(N+1)}$, so here we must take into account not the
mass $m$ itself but the product $Nm$ in the exponent. As we
discussed at the end of \Sref{concepts} the number $N$ is expected
to be large since otherwise the scalar field will influence the BH
thus violating the quasiclassical approach. So we conclude that the
decomposition \eref{sigma ser} is applicable at the whole range
$m\leq\omega\leq M$ but except the case $m=0$.

The second step is decomposition of the square root term in the
integrand from \eref{S}. Expanding it into series with respect to
$\omega$ one meets the problem at the upper bound of the integral
since $M\leq1$, so we use the following trick. As
\begin{equation*}
\fl\forall\,\omega\in[m,M]\qquad1+16M^2\omega^2>16M^2m^2\qquad\Rightarrow\qquad\left(1+4M\omega\right)^2>8M\omega+16M^2m^2,
\end{equation*}
that allows to rewrite the square root from \eref{S} in the
following form
\begin{eqnarray}\label{root ser}
\fl\sqrt{1+16M^2\left(\omega^2-m^2\right)}&=\left(1+4M\omega\right)
\sqrt{1-\frac{8M\omega+16M^2m^2}{\left(1+4M\omega\right)^2}}\nonumber\\
&=\left(1+4M\omega\right)\left\{1-\frac{4M\omega+8M^2m^2}{\left(1+4M\omega\right)^2}
+\mathcal{O}\left[\frac{\left(M\omega+2Mm^2\right)^2}{\left(1+4M\omega\right)^4}\right]\right\}\nonumber\\
&\approx4M\omega+\frac{1-8M^2m^2}{1+4M\omega}.
\end{eqnarray}

Estimating the error for \eref{root ser} one can notice that the
expression in the square brackets increases with $\omega$
decreasing. As a result the error of the decomposition applied will
be of order
$\mathcal{O}\left[M^2m^2\left(1+2m\right)^2\left(1+4Mm\right)^{-4}\right]$
and thus is small due to \eref{m bounds}.

Substituting \eref{sigma ser} and \eref{root ser} to \eref{S} we
obtain
\begin{eqnarray}\label{S final}
\frac{S\left(N,M,m\right)}{S_{\rm
BH}}\approx-\frac{1}{24\pi^3}\left.\left(\sum_{n=1}^{N}\alpha_n-N\alpha_N\right)\right|_{\omega=m}^{\omega=M},
\end{eqnarray}
where
\begin{eqnarray}\label{alpha}
\alpha_n=&\zeta^{2n}\Biggl[\frac{1+4M\omega+8M^2\left(2\omega^2-m^2\right)
+32M^3\omega\left(\omega^2-m^2\right)}{n}\nonumber\\
&+\frac{1+6M\omega+8M^2\left(2\omega^2-m^2\right)}{\pi n^2}
+\frac{3+16M\omega}{4\pi^2n^3}+\frac{1}{2\pi^3n^4}\Biggr]\nonumber\\
&+\left(1-8M^2m^2\right)\left(2\pi-1/n\right)\rme^{2\pi n}{\rm
Ei}\left[-2\pi n\left(1+4M\omega\right)\right],
\end{eqnarray}
where ${\rm Ei}(x)=\int_{-\infty}^x\rme^{-t}/t{\rm d}t$.

Expressions \eref{S final} and \eref{alpha} give an approximate
estimation for the scalar field entanglement entropy for small BH.

From \eref{S final} it is hard to achieve the power law since in any
order the corresponding terms will vanish after substituting the
integral boundaries. We have no explanation of this fact but except
that for the BH of the (sub)planck mass scale the radiation spectrum
should change to take into account quantum gravity effects. Quite
similar conclusion was made in \cite{sl10} with the help of loop
quantum gravity also.

The term proportional to logarithm of $S_{\rm BH}$ may be obtained
in the following way. As one can see, the leading-order term from
\eref{alpha} is proportional to $\zeta^{2n}/n$. Neglecting with the
higher powers of $n$ we obtain from \eref{S final} and \eref{alpha}
that
\begin{equation*}
\frac{S\left(N,M,m\right)}{S_{\rm
BH}}\propto-\left.\left(\sum_{n=1}^{N}\frac{\zeta^{2n}}{n}-\zeta^{2N}\right)\right|_{\omega=m}^{\omega=M},
\end{equation*}
that after setting $N\to\infty$ and applying \eref{zeta} transforms
to
\begin{equation}\label{S log}
\frac{S\left(\infty,M,m\right)}{S_{\rm
BH}}\propto\ln\frac{1-e^{-8\pi M^2}}{1-e^{-8\pi Mm}}.
\end{equation}
Taking into account \eref{m bounds} one can notice that $Mm\ll1$ and
then easily extract the term proportional to $\ln M$ from \eref{S
log}. However one should keep in mind that the entanglement entropy
is expressed in the units $S_{\rm BH}$ already. So the derived term
is not the logarithm correction in its common sense, but the one
proportional to it.

As we see from \eref{S final} the terms containing higher powers of
$1/n$ transform to polylogarithms of order 2, 3 and 4. So we have
obtained the other corrections to the scalar entanglement entropy.

Finally we would like to give an upper bound for the scalar
entanglement entropy. As one can notice from \eref{S} the integrand
is non-negative for any values of $\omega$ and $N$. Therefore from
\eref{S final} it follows that the entropy takes maximum values at
its boundary:
\begin{eqnarray*}
\frac{S\left(N,M,m\right)}{S_{\rm
BH}}\leq\frac{S\left(N\to\infty,1,0\right)}{4\pi}\approx1.462\cdot10^{-3}.
\end{eqnarray*}
Therefore the scalar field entanglement entropy can not be
responsible for all the entropy generation: its contribution is less
than 1\%.

\section{Discussion}\label{discussion}
Summing up we estimated the radiation entropy of the scalar field
generated with the horizon of BH with mass $M\leq1$. This paper
complements \cite{bt09} where the case $M>1$ was considered. It is
based on the similar principles.

In the paper we considered the influence of the dimension number $N$
of the in(out)side Hilbert subspace with respect to the BH horizon.
$N$ is usually taken to be infinite for simplicity, but here we
could not do so due to the smallness of BH. Estimation of the
magnitude of $N$ deserves further research.

The results demonstrate no area law dependence. We suppose this as a
consequence of quantum gravity effects. The term proportional to
logarithm of BH area is obtained. It is not the logarithm correction
term in the common sense: it is the product of logarithm and $S_{\rm
BH}$ itself. The other correction terms contain polylogarithms of
order from 2 to 4.

The upper bound of the scalar entanglement entropy does not exceed
1\% of $S_{\rm BH}$. Comparing the entanglement entropy upper bound
to \cite{bt09} we notice that its contribution is almost the same.
Such a result follows from the fact that for the upper bound
estimation on the entropy the similar assumptions were used (i.e.
$N\to\infty$). Taking the finiteness of $N$ will reduce the scalar
entropy contribution to $S_{\rm BH}$ even more but will not change
the result significantly. So we conclude that the scalar field
entanglement entropy does not dominate in $S_{\rm BH}$.

Smallness of the scalar field contribution to BH entropy might be a
consequence of our restrictions to the scalar field only. BH entropy
is detected via the particles radiated away. So the degrees of
freedom encoded with other quantum numbers should play more
significant role. Here the analogy with some register may be
observed. One-symbol language provides linear growth of the number
of possible states of the register, while even the binary language
provides the exponential one. In such a case other quantum numbers
such as spin and its projections are expected to increase the
entanglement entropy contribution significantly enough. Such a
supposition needs further investigation.

Contribution of the scalar entanglement entropy appeared to be small
comparatively to the results obtained in \cite{sre93, dss08, msk98}.
Such a discrepancy should have been expected due to the differences
in the models and assumptions considered.

Our approach has much in common with the one presented in
\cite{ilv02, ilv04} where the upper bound on the entropy is derived.
Compared to the papers here we present the analytical expression for
the entropy for small BH taking into account its dependence on $N$.

As is well-known, some of the LHC detectors are designed to explore
`new physics'; it implies looking for the additional dimensions. In
case there are such, the BH might appear during the collisions and
therefore might be detected via its Hawking radiation. The paper may
be helpful for searching for possible BH generation during
collisions. Despite here we considered the case of $3+1$-dimensional
BH, which can not be observed on the LHC, the manuscript may shed
some light on the topic since such BH is expected to be small: its
mass can not exceed 1 due to the energy restrictions. Also the
presented results can be helpful for the analysis of quark-gluon bag
model or similar ones.

\ack We are thankful to Belokolos E D for all the valuable
discussions and remarks during preparation of this work.

\bibliography{BHees}

\end{document}